# Quantifiable example of complementarity relation between optical orbital angular momentum and angular position

**HSIAO-CHIH HUANG**[1,2,*]

[1]*Institute of Atomic and Molecular Sciences, Academia Sinica, Taipei 106, Taiwan*
[2]*HC Photonics Corporation, No. 2, Technology Rd. V, Hsinchu 300, Taiwan*
*\*d93222016@ntu.edu.tw*

**Abstract:** A light beam with phase singularity (PS) characterized with azimuthally symmetric angular positions (APs) can be constructed by the rotationally symmetric superposition of *n* (*n* ∈ *N*) fractional vortex light beams with identical charges. The profile of this light beam deforms after propagation owing to its orbital angular momentum (OAM) noneigenvalue, and the deformation degree can be evaluated by the degree of phase dislocation associated with the PS. The expectation value of the mean deviation of its OAM from its characteristic charge value and the variation of the degree of phase dislocation obey a proportional relation. This light beam offers a quantifiable example of a complementarity relation between the observers of optical OAM and AP, which are the OAM noneigenvalue and intensity AP variation, respectively.

**References and links**

1. L. Allen, M. W. Beijersbergen, R. J. C. Spreeuw, and J. P. Woerdman, "Orbital angular-momentum of light and the transformation of Laguerre-Gaussian laser modes," Phys Rev A **45**, 8185-8189 (1992).35
2. A. Mair, A. Vaziri, G. Weihs, and A. Zeilinger, "Entanglement of the orbital angular momentum states of photons," Nature **412**, 313-316 (2001).37
3. H. D. Pires, H. C. B. Florijn, and M. P. van Exter, "Measurement of the spiral spectrum of entangled two-photon states," Phys. Rev. Lett. **104**, 020505 (2010).32
4. H. D. Pires, J. Woudenberg, and M. P. van Exter, "Measurement of the orbital angular momentum spectrum of partially coherent beams," Optics Letters **35**, 889-891 (2010).126
5. M. W. Beijersbergen, R. P. C. Coerwinkel, M. Kristensen, and J. P. Woerdman, "Helical-wavefront laser beams produced with a spiral phaseplate," Optics Communications **112**, 321-327 (1994).334
6. G. A. Turnbull, D. A. Robertson, G. M. Smith, L. Allen, and M. J. Padgett, "The generation of free-space Laguerre-Gaussian modes at millimetre-wave frequencies by use of a spiral phaseplate," Optics Communications **127**, 183-188 (1996).335
7. S. S. R. Oemrawsingh, E. R. Eliel, J. P. Woerdman, E. J. K. Verstegen, J. G. Kloosterboer, and G. W. Hooft, "Half-integral spiral phase plates for optical wavelengths," J. Opt. A: Pure. Appl. Opt. **6**, S288-S290 (2004).53
8. N. R. Heckenberg, R. McDuff, C. P. Smith, H. Rubinsztein-Dunlop, and M. J. Wegener, "Laser beams with phase singularities," Opt. Quan. Elec. **24**, S951-S962 (1992).336
9. I. V. Basistiy, M. S. Soskin, and M. V. Vasnetsov, "Optical wave-front dislocations and their properties," Optics Communications **119**, 604-612 (1995).56
10. M. V. Berry, "Optical vortices evolving from helicoidal integer and fractional phase steps," J. Opt. A: Pure. Appl. Opt. **6**, 259-268 (2004).5
11. J. B. Gotte, S. Franke-Arnold, R. Zambrini, and S. M. Barnett, "Quantum formulation of fractional orbital angular momentum," J. Mod. Opt. **54**, 1723-1738 (2007).48
12. J. Leach, E. Yao, and M. J. Padgett, "Observation of the vortex structure of a non-integer vortex beam," New J. Phys. **6**, 71 (2004).7
13. J. B. Gotte, K. O'Holleran, D. Preece, F. Flossmann, S. Franke-Arnold, S. M. Barnett, and M. J. Padgett, "Light beams with fractional orbital angular momentum and their vortex structure," Opt. Exp. **16**, 993-1006 (2008).1
14. N. Bohr, *Atomic Theory and the Description of Nature* (Cambridge University Press, Cambridge UK, 1934).423

15. E. Yao, S. Franke-Arnold, J. Courtial, S. Barnett, and M. Padgett, "Fourier relationship between angular position and optical orbital angular momentum," Opt. Exp. **14**, 9071-9076 (2006).398
16. S. Franke-Arnold, S. M. Barnett, E. Yao, J. Leach, J. Courtial, and M. Padgett, "Uncertainty principle for angular position and angular momentum," New J. Phys. **6**, 103 (2004).52
17. M. V. Berry, "Quantal Phase-Factors Accompanying Adiabatic Changes," Proc. R. Soc. Lond. A **392**, 45 (1984).128
18. J. Leach, M. J. Padgett, S. M. Barnett, S. Franke-Arnold, and J. Courtial, "Measuring the orbital angular momentum of a single photon," Phys. Rev. Lett. **88**, 257901 (2002).154
19. B. E. A. Saleh and M. C. Teich, *Fundamentals of Photonics* (Wiley, New York, 1991), Vol. Chap. 4.357
20. H.-C. Huang, Y.-T. Lin, and M.-F. Shih, "Measuring the fractional orbital angular momentum of a vortex light beam by cascaded Mach-Zehnder interferometers," Optics Communications **285**, 383-388 (2011).157
21. H.-C. Huang, "Various angle periods of parabolic coincidence fringes in violating Bell inequality with high-dimensional two-photon entanglement," arXiv **1804.08930**(2018).424
22. G. Stephenson and P. M. Radmore, *Advanced Mathematical Methods for Engineering and Science Students* (Cambridge University Press, Cambridge, 1993), Vol. Chap. 5.354

---

### 1. Introduction

An optical vortex (OV) with azimuthal phase structure $e^{im\phi}$ can carry an OAM eigenvalue $m\hbar$ per photon and its wavefront is helical around its own propagation axis, where the integer $m$ is the topological charge and $\phi$ is the azimuthal coordinate [1]. The natural existence of numerous OAM eigenvalues has been experimentally proven in previous studies using incoherent photons [2, 3] and a partial coherent light beam [4]. This eigenvalue characteristic represents the exact matching between the phase shift of a complete cycle of a wave $2\pi$ and the positive integer division of the phase shift of the helical wavefront per cycle. However, the arbitrary phase shift of helical wavefront per cycle can be generated artificially by means of a spiral phase plate [5-7] and folk hologram [8, 9]. In the case of a non-integer multiple of $2\pi$, there is a phase dislocation of the helical wavefront by definition. A light beam in this case is called a fractional vortex (FV) in this report and has an azimuthal phase structure $e^{iM\phi}$, where its characteristic charge $M = m + \mu$ and $\mu$ is an improper fraction [10].

The quantum state of an FV is the superposition of numerous optical OAM eigenmodes with a functional weight [10, 11] and it carries an OAM noneigenvalue. As well as the functional weight, the departure of the OAM noneigenvalue from $M\hbar$ is a sinusoidal function of $M\hbar$ [10-12]. The phase dislocation can be referred to that the PS is characterized with one of the angular positions (APs) of an FV. In its near-field image, the beam shape is almost round and a low intensity line exists along that AP, but in the far-field, these almost round and linear shapes no longer exist [10, 12, 13] owing to the diffraction with various Guoy phases from the composed OAM eigenmodes [13]. The degree of phase dislocation is an important factor because it indicates the degree of beam deformation [13] and results in a variety of localized vortices [12]. This beam deformation can be illustrated from the perspective of OAM; the arbitrarily well-defined beam profiles in the near-field that carry the OAM noneigenvalue will deform azimuthally after propagation, whereas a light beam carrying an OAM eigenmode always has a stable azimuthal structure in free space.

An object has pairs of complementary properties that cannot all be observed simultaneously [14]. This rule holds by the complementarity principle, and each of the pairs is mathematically conjugated. Optical OAM and AP are two observables of a conjugate pair [15, 16]. However, the sinusoidal function of the departure of the OAM noneigenvalue from $M\hbar$ for a light beam with PS characterized with an AP has a certain amplitude. One may inquire whether light beams with PS characterized with $n$ APs have a correlation between the OAM noneigenvalue and the image intensity AP.

In this article, it is shown that a structured light beam with PS characterized with $n$ arbitrary APs can be constructed by the rotational superposition of $n$ FVs with numerous phase shifts. As with the case of the FV, there are low-intensity linear-shapes along with the $n$

APs in the near-field image of this structured light beam, and these shapes no longer exist in the far-field image. When the superposition has rotational symmetry and there are no phase shifts between the FVs, the PS is characterized with the azimuthally symmetric APs of the structured light beam. This structured light beam is called a multiple fractional vortex (MFV) and is also denoted as MFV$n$ in this report. The mean deviation of the OAM from $M\hbar$ and the variation of the degree of phase dislocation with respect to $M$ for an MFV$n$ are both sinusoidal functions of $M$ with quantized amplitudes inversely and proportionally related to $n$ respectively; their amplitudes result in an inverse proportion. These inverse and proportional amplitudes can be illustrated by observing the OAM and AP, respectively. This inversely proportional relation between the mean deviation of the OAM and the variation of phase dislocation degree can readily be regarded as a quantifiable example of the complementarity relation between the two observables of OAM and AP. Thus, a light beam with PS that is characterized with azimuthally symmetric APs offers a quantifiable example of the complementarity relation between the mean deviation of the OAM from its characteristic charge and the variation of its degree of phase dislocation. The remaining content is organized as follows. Section 2 introduces MFV$n$s. Section 3 outlines their experimental construction and verifies their rotational symmetry. Section 4 illustrates the inversely proportional relation and its physical meaning. Section 5 presents concluding remarks.

## 2. Light beam with PS characterized with arbitrary APs

An arbitrary light beam can be divided into two beams that are identical in power. These two beams can then travel different path lengths and subsequently overlap completely to form a single beam. This is the primary function of a Mach–Zehnder interferometer (MZI). If the light beam is coherent, the overlap causes superposition to occur, and the path length difference is equivalent to the phase shift $\delta$ between the divided beams. Thus, the power changes without changing the field structure between the light beams before division and after superposition with $\delta$ between the two divided beams. The power change can be illustrated by multiplying a term $e^{-i\delta}$ by the complex electric field of one of the two divided beams. A relative rotation angle $\theta$ can be introduced to the two divided beams by an MZI with geometric phase $\theta$ (or Berry phase [17], equals to the relative rotation angle in units of radians). If an OV is input to an MZI with $\theta$, then $\theta$ will cause various phase shifts $m\theta$ in addition to $\delta$ between the two divided beams in correspondence with the various charges of $m$ [18]. An OV input into an MZI with $\theta$ yields results similar to those obtained by inputting an arbitrary coherent light beam into an MZI; the power changes without changing in the field structure or the azimuthal part of the field structure between the OVs before division and after superposition with $\delta$ and $m\theta$. However, the field structure does not remain unchanged if an FV is input into an MZI with $\theta$.

A field state is the quantum state of a light beam in quantum theory. A characterized AP $\alpha$ of the PS can be incorporated into the quantum state of an FV by using $|M(\alpha)\rangle$ [11], which is introduced in Appendix A. If an FV is input into an MZI with $\theta$, the unnormalized quantum state after superposition with $\delta$ and relative rotation angle $\theta$ is the original quantum state plus a rotation operator $\hat{U}$ acting on the original quantum state as $|M(\alpha)\rangle + e^{-i\delta}\hat{U}(\theta)|M(\alpha)\rangle$, where $\hat{U}$ is also introduced in Appendix A. The azimuthal structure of the latter term is not generally equal to that of the former term (they are equal only if $\mu = 0$ and $|M(\alpha)\rangle = |m\rangle$) owing to the new characterized AP of the PS as $\hat{U}(\theta)|M(\alpha)\rangle = e^{-im\theta}|M(\alpha \oplus \theta)\rangle$ (cf. Eq. (A7)), and the quantum state after superposition is $|M(\alpha)\rangle + e^{-i(m\theta+\delta)}|M(\alpha \oplus \theta)\rangle$. Consequently, there are two characterized APs, $\alpha$ and $\alpha + \theta$, of the PS in the superposed light

beam. Similarly, a light beam with the PS characterized with $n$ arbitrary APs can be constructed by performing $n - 1$ divisions and $n - 1$ overlaps.

To simulate the beam pattern for FVs, the wave function $\Psi(\rho,\phi,z=0) = e^{-\rho^2/w^2} e^{iM\phi}$ is considered [10, 11], where $w$ is the beam waist radius and $\rho$ is the radial coordinate. The propagation solution for the wave function $\Psi(\rho,\phi,z)$ can be evaluated through a transfer function in free space $e^{i\sqrt{k^2-\kappa^2}z}$ [19], where $k$ and $\kappa$ are the wave number and transverse wave number, respectively. Figure 1(a) presents the near-field intensity images of a FV with $M = 2/3$. Figure 1(b) presents two near-field intensity images of the superposition of two FVs with identical charges of 2/3. Each of the images has the PS characterized with two APs. Except in the regions with the PS in each, the two images are azimuthally isotropic in terms of intensity according to the relation $M(\pi-\theta)=\delta$, where $\theta$ and $\delta$ are compared between two images in Fig. 1(a). The left image in Fig. 1(b) has $\theta = 180°$ and $\delta = 0$, and the right image has $\theta = 90°$ and $\delta = \pi/3$. In the left image in Fig. 1(b), the two APs are located symmetrically with respect to their own propagation axis, whereas those in the right image are located asymmetrically with respect to their own propagation axis. Figure 1(c) depicts three near-field intensity images of the superposition of four FVs with identical charges of 2/3. Each of the images includes the PS characterized with four APs. Except in the characterized AP regions in each, the left and middle images in Fig. 1(c) and the right image in Fig. 1(c) are azimuthally isotropic in intensity according to the relation $M(\pi/2-\theta)=\delta$, in which $\theta$ and $\delta$ are compared between two images in the left of Fig. 1(b) and between two images in the right of Fig. 1(b), respectively. For the left and middle in Fig. 1(c), $\theta = 90°$ and $\delta = 0$ and $\theta = 45°$ and $\delta = \pi/6$, respectively, and $\theta = 45°$ and $\delta = \pi/6$ for the right image in Fig. 1(c). In the left image in Fig. 1(c), the four APs are located symmetrically with respect to their own propagation axis, whereas those in the middle and right images are located asymmetrically with respect to their own propagation axis. All of the light beams in Fig. 1 can be obtained experimentally by using cascaded MZIs with various geometric phases and various phase shifts. Similarly, by cascaded MZIs, a light beam composed of $2^t$ superimposed FVs, $t = 1, 2, 3, ...$ can be azimuthally isotropic in intensity (except when there is the PS characterized with $2^t$ APs in each) according to the relation $M(2\pi/2^t - \theta_t) = \delta_t$, $M \notin Z$.

Similarly, a light beam composed of $n \neq 2^t$ superimposed FVs can be azimuthally isotropic in intensity (except when there is PS characterized with $n$ APs in each) according to the relations among $M$, $n$, $\theta$, and $\delta$.

The characterized APs of the PS divide the near-field image into $n$ pieces. The images in the symmetric cases of $n = 1, 2,$ and $4$ in Fig. 1(a) and on the left sides of Figs. 1(b) and 1(c) have equal divisions into $n$ pieces, whereas those in the asymmetric cases do not. A light beam with the PS characterized with $n$ symmetric APs is an MFV$n$, as mentioned in Section 1. From here onward, an MFV$n$ with $M$ represents a light beam composed of $n$ superimposed FVs with $n$ identical charges of $M$, $\theta = \theta_n = 2\pi/n$, and $\delta = 0$.

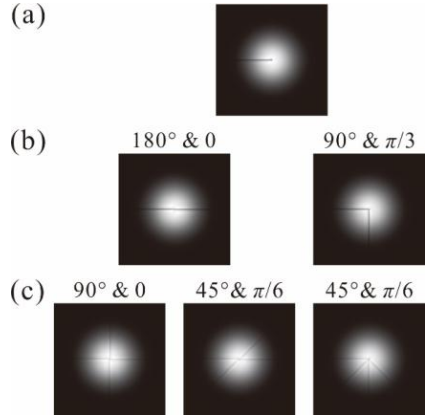

Fig. 1. Light beam with the PS characterized with arbitrary APs. (a) Simulated intensity image of an FV with $M = 2/3$. (b) Left: simulated intensity image with the superposition of two FVs with two identical charges of $M = 2/3$, $\theta = 180°$, and $\delta = 0$ (MFV2). Right: simulated intensity image with the superposition of two FVs with two identical charges of $M = 2/3$, $\theta = 90°$, and $\delta = \pi/3$. (c) Left: simulated intensity image with the superposition of two light beams from the left of Fig. 1(b) with $\theta = 90°$ and $\delta = 0$ (MFV4, four FVs with four identical charges of $M = 2/3$, $\theta = 90°$, and $\delta = 0$). Middle: simulated intensity image with the superposition of two light beams from the left of Fig. 1(b) with $\theta = 45°$ and $\delta = \pi/6$. Right: simulated intensity image with the superposition of two light beams from the right of Fig. 1(a) with $\theta = 45°$ and $\delta = \pi/6$.

There are two variables for a light beam with the PS characterized with $n$ APs—specifically, the number $n$ and the characterized APs for the PS. The characterized APs for the PS in asymmetric case are the parts of those in the symmetric case with a larger number of $n$. For example, the two APs of the phase dislocation in the right image in Fig. 1(b) are two of the four APs in the left image in Fig. 1(c), and similarly for the two images in the middle and right of Fig. 1(c). In particular, the symmetric cases are compared herein to illustrate the underlying physical relation between the AP and OAM observed in light beams with the PS characterized with $n$ APs. In fact, the symmetric cases are just the cases required for comparison to elucidate this underlying physical relation. This comparison with only the symmetric cases in elucidating the underlying physical relation also can be understood from the OAM eigenmodes. A symmetric case has $\theta = \theta_n = 2\pi/n$ and $\delta = 0$, and its nonvanishing OAM eigenmodes are $m' \bmod n = 0$ owing to the completely constructive interference [18]. In contrast, in asymmetric cases, the OAM eigenmodes interfere partially destructively according to their topological charges and nonzero phase shifts of their overlaps. A large number of superpositions implies significant destructive interference. To elucidate the underlying physical relation based on the comparison of the OAM eigenmodes and their weights from the destructive interference, a smaller number of superpositions in an asymmetric case therefore can be included in a larger number of superpositions in a symmetric case.

Figure 2(a) presents near-field phase profiles of an FV, MFV2, and MFV4 with $M = 2/3$. The phase gradients with respect to $\phi$ are identical for all of the helical wavefronts between the FV, MFV2, and MFV4, because they are superposed with identical charges. According to Fig. 2(a), the phase variations in each of the sections are 2.09, 1.05, and 0.52 rad for the FV, MFV2, and MFV4, respectively. A large $n$ implies a small phase variation in each section. The number of superpositions can be applied for arbitrary $n$, and the phase variation is $2\pi M/n$ in each of the sections of MFV$n$. Owing to the rotational symmetry superposition, MFV$n$ has $C_n$ rotational symmetry in the intensity image and phase profile, as in the cases of 1-, 2-, and 4-fold rotation symmetries that are depicted in Figs. 1 and 2(a). Figure 2(b) shows far-field intensity images of the FV, MFV2, and MFV4 with $M = 2/3$. Although these images are not round and the lineal PS no longer exists, they are the inevitable outcomes of near-field intensity images with spatial evolution. The characteristics of the 1-, 2-, and 4-fold rotational symmetries in these far-field images provided in Fig. 2(b) are due to these inevitable outcomes.

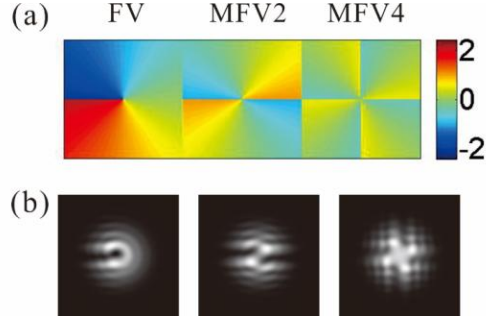

Fig. 2. Phase profile and far-field intensity images of MFV$n$. (a) Three simulated phase profiles of an FV, MFV2, and MFV4 with identical charges $M = 2/3$. The values in the color bar have units of radians, and the full ranges of the phase variation in each of the sections are 2.09, 1.05, and 0.52 rad for the FV, MFV2, and MFV4, respectively. (b) Three far-field intensity images of the FV, MFV2, and MFV4 with identical charges $M = 2/3$.

## 3. Experimental construction and verification of rotational symmetry of MFV$n$

The setup of an MZI with $\theta$ is shown in Fig. 3(a). The first-order light beam from a hologram is divided into two by the first beam splitter. Each of the two beams passes through a mirror and a Dove prism. The rotation angle $\theta$ between the light beams in arms 1 and 2 is constructed by the rotation angle $\theta/2$ between the two Dove prisms [18]. Then, the two beams are overlapped by the second beam splitter. Two superposed light beams with zero and $\pi$ phase shifts are obtained in turn by a piezo stage in one of the two exit arms. For the interference in the ideal wavefront distribution, an imaging lens is used [20], and the image is recorded using a charge coupled device (CCD) camera.

In Fig. 3(b), MFV2 and MFV4 are constructed using MZIs with $\theta = \pi$ and $\theta = \pi/2$, respectively. An FV is input into the MZI with $\theta = \pi$, and the output is MFV2, as shown in the upper part of Fig. 3(b). MFV4 is constructed by cascading the second MZI with $\theta = \pi/2$, following the first MZI, as shown in the lower part of Fig. 3(b). MFV$n$ can be constructed in a similar manner.

That MFV$n$ has rotational symmetry $C_n$ was proven theoretically in a previous study based on its quantum state [21]. The $C_n$ rotational symmetry of MFV$n$ can be experimentally verified by using the interferometers depicted in Fig. 3(c). In the upper part of Fig. 3(c), MFV2 is input into an MZI with $\theta = \pi$, and the zero and $\pi$ phase shifts output at ports A and B are completely constructive and destructive images of MFV2, respectively. This outcome proves that the two pieces of MFV2 are identical, and MFV2 has rotational symmetry $C_2$. Similarly, the four pieces of MFV4 are identical, as verified by an MFV with $\theta = \pi/2$ in the lower part of Fig. 3(c). Despite the unstable interference outcome on the optical wavelength scale, MFV can be generated identically by holography. In Fig. 3(d), the left part shows a 4-fork hologram to generate MFV4, and the right part present the first-order image from the 4-fork hologram and the verification of MFV4 by an MZI with $\theta = \pi/2$.

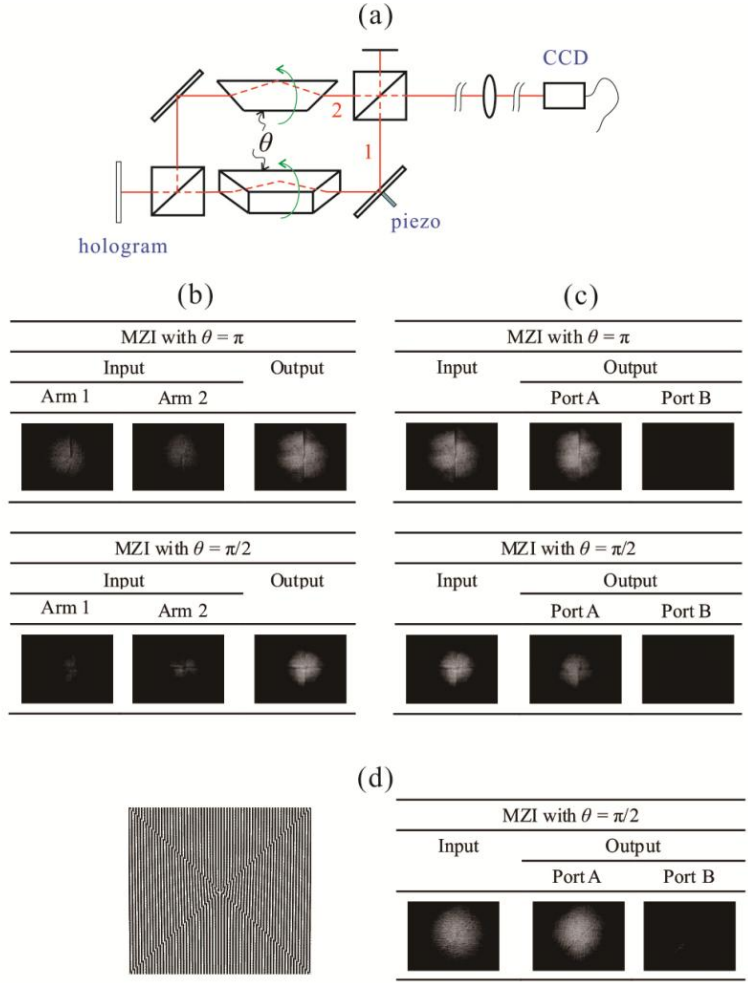

Fig. 3. Construction and rotational symmetry verification of MFV*n*. (a) MZI with *θ* and imaging detection. (b) Constructions of MFV2 and MFV4 are shown in the upper and lower tables for MZIs with *θ* = π and *θ* = π/2, respectively. (c) Verifications for MFV2 and MFV4 are shown in the upper and lower tables for MZIs with *θ* = π and *θ* = π/2, respectively. (d) Four-fork hologram used to generate MFV4, generated image, and verification of MFV4 by using an MZI with *θ* = π/2. *M* = 1/3 is taken as an example in (b), (c), and (d).

## 4. Relation between OAM mean and variation of phase dislocation degree

The OAM probability distribution and OAM mean of the FV were evaluated [11] in a previous study as $P_{m'}(M) = \sin^2(\mu\pi)/\left[\pi^2(M-m')^2\right]$ (cf. Eq. (A5)) and $\overline{M} = M - \sin(2M\pi)/2\pi$ (cf. Eq. (A6)), respectively, where $m'\hbar$ is the OAM eigenvalue and $\overline{M}$ has units of $\hbar$. The quantum state of MFV*n* is

$$|Mn\rangle = \frac{|Mn'\rangle}{\sqrt{\langle Mn'|Mn'\rangle}}, \quad |Mn'\rangle = \sum_{k=0}^{n-1} \hat{U}\left(2\pi \times \frac{k}{n}\right)|M\rangle, \qquad (1)$$

where $|M\rangle$ is the quantum state of the FV and $\hat{U}$ is the rotation operator, as introduced in Appendix A. According to Eq. (1), the probabilities for the decomposition of MFV2 into integer OAM modes can be evaluated using (cf. Eq. (B3))

$$P_{m'}[M2(M)] = |\langle m'|M2\rangle|^2 = \frac{2}{\pi^2}\sin^2\left(\frac{M\pi}{2}\right) \times \frac{1+\cos(\pi m')}{(M-m')^2}. \quad (2)$$

Owing to the completeness of the OAM basis state, the probabilities add up to unity $\sum_{m'=-\infty}^{\infty} P_{m'}(M2) = 1$ (cf. Eq. (B4)). The mean OAM of MFV2 is given by $\overline{M2} = \sum_{m'=-\infty}^{\infty} m' P_{m'}(M2) = M - \sin(M\pi)/\pi$ (cf. Eq. (B5)). According to Eq. (1), the probabilities for the decomposition of MFV4 into integer OAM modes can be evaluated using (cf. Eq. (B10))

$$P_{m'}[M4(M)] = |\langle m'|M4\rangle|^2 = \frac{4}{\pi^2}\sin^2\left(\frac{M\pi}{4}\right) \times \frac{1+\cos(\pi m')+2\cos(\frac{\pi}{2}m')}{(M-m')^2}. \quad (3)$$

Owing to the completeness of the OAM basis state, the probabilities add up to unity $\sum_{m'=-\infty}^{\infty} P_{m'}(M4) = 1$ (cf. Eq. (B11)). The OAM mean of MFV4 is given by $\overline{M4} = \sum_{m'=-\infty}^{\infty} m' P_{m'}(M4) = M - \sin(M\pi/2)/(\pi/2)$ (cf. Eq. (B12)). According to Eqs. (A6), (B5), and (B12), the OAM mean of MFVn should be evaluated using

$$\overline{Mn}(M) = M - \frac{n}{2\pi}\sin\left(M\frac{2\pi}{n}\right). \quad (4)$$

Because the quantum state of MFVn is not an OAM eigenstate, its mean OAM is not necessarily equal to the product of $M$ and $\hbar$. However, the rotational symmetry superposition and $M$ jointly regulate the sinusoidal relations between the deviation of the OAM mean from the characteristic value $M\hbar$ and $M$, and the amplitude of these sinusoidal functions is proportional to the number of rotational symmetry superpositions $n$. By using Eq. (4), the relations between $\overline{Mn} - M\hbar$ and $M$ were obtained for the FV, MFV2, and MFV4 and are shown by the blue curves in Fig. 4(a). They are sinusoidal, and their amplitudes vary in increments of $n\hbar/2\pi$, as indicated by the green markings.

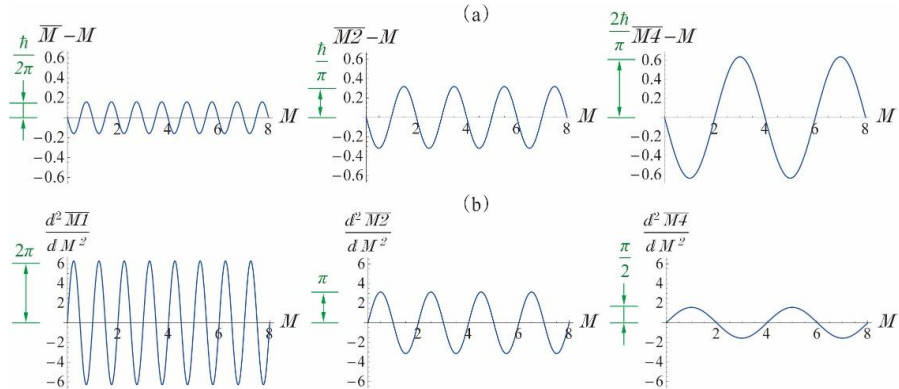

Fig. 4. Proportionality of the mean deviation of the OAM from $M$ and the variation of the phase dislocation degree with respect to $M$. (a) Periodic functions of $M$ showing the difference between $\overline{M}$ and $M\hbar$ for an FV, MFV2, and MFV4 (blue curves; cf. Eq. (4)), where green markings indicate the amplitudes of these periodic functions for $\hbar/2\pi$, $\hbar/\pi$, and $2\hbar/\pi$,

respectively. (b) Periodic functions of *M* representing the second derivatives of $\overline{M}$ with respect to *M* for the FV, MFV2, and MFV4 (blue curves; cf. Eq. (6)), where the green markings indicate the amplitudes of these periodic functions for $2\pi$, $\pi$, and $\pi/2$, respectively. The products of the two amplitudes are $\hbar$ for the FV, MFV2, and MFV4.

*M* indicates the degree of phase dislocation for a light beam with the PS characterized with one AP (FV), but does not do so for those with the PS characterized with *n* APs. However, the variation of the OAM mean with respect to the charge *M*, $d\overline{Mn}/dM$, does. From Eq. (4),

$$\frac{d\overline{Mn}}{dM} = 1 - \cos\left(M\frac{2\pi}{n}\right). \tag{5}$$

Eq. (5) is a sinusoidal function of *M*, whose amplitude is between zero and two and whose period is proportional to *n*. Its maximum of two corresponds to the maximum phase difference $\pi$ between the two sides of the phase dislocation, its minimum of zero corresponds to the zero phase difference, and other values correspond to the intermediate phase differences in monotonous variations. According to Eq. (5), the maximum degree of phase dislocation occurs at $\left(M - \frac{n}{2}\right) \bmod n = 0$.

Franke-Arnold *et al.* reported on the uncertainty principle of the OAM and AP, the conjugate variable of the OAM [15], for a light beam that lacks an angular component, or a sector light beam [16]. A sector light beam has the OAM uncertainty and the AP uncertainty. The OAM uncertainty presents the range of the OAM spectrum, whereas the AP uncertainty presents the range of the intensity distribution of the azimuthal coordinate. Although there is a high probability weight in a range of OAM eigenmodes [11] and low intensity for a range of APs [10], the uncertainties of the OAM and AP for an FV cannot be determined readily owing to the divergence of the OAM variance [11]. By comparing of these two ranges for various MFV*n*s, as shown in Appendix D, it is evident that a large degree of phase dislocation implies a small AP uncertainty and, simultaneously, a large OAM uncertainty. Therefore, $d\overline{Mn}/dM$, the degree of phase dislocation, can quantitatively present OAM and AP uncertainties for MFV*n* that are complementary to each other, although these two uncertainties cannot be determined readily [11]. $d\overline{Mn}/dM$ should be an important factor, because it presents OAM and AP uncertainties, indicates the degree of beam deformation, and may result in a variety of localized vortices for MFV*n*s.

$d^2\overline{Mn}/dM^2$ is the variation of the degree of phase dislocation with respect to the charge for MFV*n*. From Eq. (5),

$$\frac{d^2\overline{Mn}}{dM^2} = \frac{2\pi}{n}\sin\left(M\frac{2\pi}{n}\right). \tag{6}$$

$d^2\overline{Mn}/dM^2$ is a sinusoidal function, and its amplitude varies in increments of $2\pi/n$. By using Eq. (6), the relationships between $d^2\overline{Mn}/dM^2$ and *M* for the FV, MFV2, and MFV4 were obtained and are shown by the blue curves in Fig. 4(b). The quantized amplitudes are indicated by the green markings. The periods of $\overline{Mn} - M\hbar$ and $d^2\overline{Mn}/dM^2$ are identical and proportional to *n*; however, their amplitudes are inversely proportional with proportionality constant $\hbar$. Indeed, a large number of the characterized APs of the PS implies a large expectation value of the mean deviation of the OAM from $M\hbar$ and, simultaneously, a small expectation value of the variation of the phase dislocation degree with respect to the charge.

This proportionality relation is a quantifiable example of the complementary relation between the OAM and AP. The OAM eigenmodes of $m' \bmod n \neq 0$ vanish owing to the completely destructive interference, and the resolution limit of the OAM spectra is $n$, as well as the interval of the OAM eigenmodes. Owing to the decrease in the OAM eigenmode resolution by $1/n$ for MFV$n$, the fluctuation of the mean OAM with respect to the characteristic value $M\hbar$, as well as the amplitude of Eq. (4), is larger by a factor of $n$ in comparison to that of an FV. In contrast, the AP range of a complete cycle $2\pi$ in MFV$n$ is used to construct the PS characterized with $n$ APs. Equivalently, the AP range is reduced to $2\pi/n$ or the phase variation is reduced to $2\pi M/n$ to construct the PS with one AP in MFV$n$. Owing to the reduction of the AP range or phase variation, the variation of the degree of PS with respect to $M$ ($d^2\overline{Mn}/dM^2$) for MFV$n$ is reduced.

## 5. Conclusion

A structured light beam with the PS characterized with azimuthally symmetric APs (MFV$n$) is constructed by the rotationally symmetric superposition of $n$ FVs with identical charges $M$. A quantifiable example of the complementarity relation between OAM and AP is exhibited by MFV$n$: a large number of characterized APs of the PS in such a light beam implies a large expectation value of the mean deviation of the OAM from $M\hbar$ and, simultaneously, a small expectation value of the variation of the phase dislocation degree with respect to the charge.

The variations of the mean deviation of the OAM from $M\hbar$ and the phase dislocation degree with respect to the charge are discrete, which are a result of the number of characterized APs of the PS. Thus, this complementarity relation between OAM and AP from MFV$n$ is discrete in variation. By contrast, the uncertainty principle between OAM and AP is continuous in variation, because the uncertainties of OAM and AP are continuous variations. MFV$n$ has the characterization in its structure; thus, MFV$n$ may be useful in a variety of scientific fields.

# Appendices

## A. Quantum state of an FV

The quantum state of an FV is denoted by $|M(\alpha)\rangle$ [11], where $M = m + \mu$ and the parameter $\alpha$, bounded by $0 \leq \alpha < 2\pi$, is the AP of the discontinuity. The following function $f_\alpha(\phi)$ can be introduced:

$$f_\alpha(\phi) = \begin{cases} 1, & 0 \leq \phi < \alpha \\ 0, & \alpha \leq \phi < 2\pi \end{cases}. \tag{A1}$$

By using Eq. (A1), the azimuthal part of a light field can be defined as

$$\langle \phi | M(\alpha) \rangle \equiv e^{im\phi} e^{i\mu[\phi + 2\pi f_\alpha(\phi) - \alpha]}. \tag{A2}$$

Based on the completeness relation and Eq. (A2), the overlap amplitude between the FV state and its rotated state with an angle $\alpha$ is

$$\langle M(0) | M(\alpha) \rangle = \frac{1}{2\pi} \int_0^{2\pi} d\phi \, e^{i2\pi\mu[f_\alpha(\phi) - f_0(\phi)]} e^{i\mu(0-\alpha)}$$
$$= \frac{1}{2\pi} \left[ \alpha e^{i(2\pi-\alpha)\mu} + (2\pi - \alpha) e^{-i\alpha\mu} \right]. \tag{A3}$$

The FV state can be decomposed into integer OAM eigenmodes, whose probability distribution can be obtained by setting $\langle M(0)| = \langle m'|$ in Eq. (A3)

$$c_{m'}[M(\alpha)] = \langle m' | M(\alpha) \rangle = \frac{ie^{i(m-m')\alpha}}{2\pi(M-m')}(1 - e^{i2\pi\mu}). \tag{A4}$$

The state depends on $\alpha$. However, the probability, the square modulus of $c_{m'}[M(\alpha)]$, is independent of $\alpha$:

$$P_{m'}(M) = |\langle m' | M(\alpha) \rangle|^2 = \frac{\sin^2(\mu\pi)}{\pi^2(M-m')^2}. \tag{A5}$$

For $M \in Z$, $P_{m'}(M) = \delta_{Mm'}$, where the case of $m' = M$ is according to L'Hôpital's rule. According to Eq. (A5), the OAM mean of an FV is

$$\overline{M} = \sum_{m'=-\infty}^{\infty} m' P_{m'}(M) = M - \frac{\sin(2M\pi)}{2\pi}, \tag{A6}$$

where Eq. (C3) has been used.

The state resulting from a unitary operator $\hat{U}(\beta)$ is the effect of the rotation for the FV state and the inclusion of a phase term $e^{-im\beta}$:

$$\hat{U}(\beta) | M(\alpha) \rangle = e^{-im\beta} | M(\alpha \oplus \beta) \rangle, \tag{A7}$$

where the parameter $\beta$ bounded by $0 \leq \beta < 2\pi$ is the action on the state and $\alpha \oplus \beta = (\alpha + \beta) \bmod 2\pi$ yields a result in the range $[0, 2\pi)$ owing to the $2\pi$ modulo. The multiplication of rotation operators has the combination characteristic

$$\hat{U}(\beta_1) \hat{U}(\beta_2) | M(\alpha) \rangle = \hat{U}(\beta_1 \oplus \beta_2) | M(\alpha) \rangle. \tag{A8}$$

Some useful formulas are as follows. By utilizing Eqs. (A3) and (A7),

$$\langle M(\alpha)|\hat{U}(\beta)|M(\alpha)\rangle = \frac{e^{-im\beta}}{2\pi}\left[\beta e^{i(2\pi-\beta)\mu} + (2\pi-\beta)e^{-i\beta\mu}\right], \quad (A9)$$

which is independent of $\alpha$. The real part of Eq. (A9) is

$$\mathrm{Re}\langle M(\alpha)|\hat{U}(\beta)|M(\alpha)\rangle = \frac{1}{2\pi}\left[\beta\cos(2\pi\mu-\beta M) + (2\pi-\beta)\cos(\beta M)\right]. \quad (A10)$$

By using Eqs. (A4) and (A7),

$$\langle m'|\hat{U}(\beta)|M(\alpha)\rangle = e^{-im\beta}\frac{ie^{i(m-m')(\beta\oplus\alpha)}}{2\pi(M-m')}(1-e^{i2\pi\mu}). \quad (A11)$$

Its square modulus is

$$\left|\langle m'|\hat{U}(\beta)|M(\alpha)\rangle\right|^2 = \frac{\sin^2(\mu\pi)}{\pi^2(M-m')^2}. \quad (A12)$$

By employing Eq. (A11),

$$\langle m'|\hat{U}(\beta_1)|M(\alpha)\rangle\langle m'|\hat{U}(\beta_2)|M(\alpha)\rangle^* = \frac{\sin^2(\pi\mu)}{\pi^2(M-m')^2}e^{im'(\beta_2-\beta_1)}, \quad (A13)$$

which is independent of $\alpha$ but dependent on the difference between $\beta_1$ and $\beta_2$.

## B. Unity summation of probability and OAM mean for $|M2\rangle$ and $|M4\rangle$

For $n=2$,

$$\langle M2'|M2'\rangle = \left[\langle M| + \langle M|\hat{U}^\dagger(\pi)\right]\left[|M\rangle + \hat{U}(\pi)|M\rangle\right]$$
$$= 2\left[\langle M|M\rangle + \mathrm{Re}\langle M|\hat{U}(\pi)|M\rangle\right] = 2\left[1+\cos(M\pi)\right], \quad (B1)$$

where the unitary property of $\hat{U}$, $\cos(2\pi\mu-\pi M) = \cos(\pi M - 2\pi m) = \cos(\pi M)$, and Eqs. (A3) and (A10) have been used.

$$\left|\langle m'|M2'\rangle\right|^2 = \left[\langle m'|M(\alpha)\rangle + \langle m'|\hat{U}(\pi)|M(\alpha)\rangle\right]\left[\langle m'|M(\alpha)\rangle + \langle m'|\hat{U}(\pi)|M(\alpha)\rangle\right]^*$$
$$= \left|\langle m'|M(\alpha)\rangle\right|^2 + \left|\langle m'|\hat{U}(\pi)|M(\alpha)\rangle\right|^2 + 2\mathrm{Re}\left[\langle m'|M(\alpha)\rangle\langle m'|\hat{U}(\pi)|M(\alpha)\rangle^*\right] \quad (B2)$$
$$= \frac{2\sin^2(\mu\pi)}{\pi^2(M-m')^2}\left[1+\cos(\pi m')\right],$$

where Eqs. (A5), (A12), and (A13) have been used. According to Eqs. (1), (B1), and (B2), the OAM probability distribution of MFV2 is

$$P_{m'}\left[M2(M)\right] = \frac{\left|\langle m'|M2'\rangle\right|^2}{\langle M2'|M2'\rangle} = \frac{2}{\pi^2}\sin^2\left(\frac{M\pi}{2}\right) \times \frac{1+\cos(\pi m')}{(M-m')^2}, \quad (B3)$$

where $\sin^2(\mu\pi) = \sin^2(M\pi)$ has been used. The unity summation of the OAM probability of $|M2\rangle$ can be proven from Eq. (B3),

$$\sum_{m'=-\infty}^{\infty} P_{m'}(M2) = \frac{2}{\pi^2}\sin^2\left(\frac{M\pi}{2}\right)\sum_{m'=-\infty}^{\infty}\frac{1+\cos(\pi m')}{(M-m')^2} = 1, \quad (B4)$$

where Eqs. (C1) and (C2) have been used. The OAM mean of MFV2 can be calculated from Eq. (B3) as

$$\sum_{m'=-\infty}^{\infty} m' P_{m'}(M2) = \frac{2}{\pi^2}\sin^2\left(\frac{M\pi}{2}\right)\sum_{m'=-\infty}^{\infty}\frac{m'+m'\cos(\pi m')}{(M-m')^2} = M - \frac{\sin(M\pi)}{\pi}, \quad (B5)$$

where Eqs. (C3) and (C4) have been used. It is interesting to consider the $M \in Z$ case. As $M \bmod 2 = 0$, Eq. (B3) becomes

$$P_{m'}[M2(M \bmod 2 = 0)] = \begin{cases} 0, & m' \neq M \\ \dfrac{\sin^2(\mu\pi)}{\pi^2(M-m')^2} = 1, & m' = M \end{cases} = \delta_{Mm'}, \quad (B6)$$

where L'Hôpital's rule is used for $m' = M$ and $\delta_{Mm'}$ is the Kronecker delta function. According to Eq. (B6), MFV2 with $M \bmod 2 = 0$ is equal to OV with $m \bmod 2 = 0$. Thus, MFV2 with $M \bmod 2 = 0$ can be generated by the rotationally symmetric superposition of two OVs with $m \bmod 2 = 0$, the case of completely constructive interference [18, 20]. As $M \bmod 2 = 1$, Eq. (B3) becomes

$$P_{m'}[M2(M \bmod 2 = 1)] = \frac{2[1+\cos(\pi m')]}{\pi^2(M-m')^2} = \begin{cases} \dfrac{2[1+\cos(\pi m')]}{\pi^2(M-m')^2}, & m' \bmod 2 = 0 \\ 0, & m' \bmod 2 = 1 \end{cases}, \quad (B7)$$

where the OAM modes of $m' \bmod 2 = 1$ vanish owing to the completely destructive interference in an MZI with $\theta = \pi$ [18, 20]. According to Eq. (B7), MFV2 with $m \bmod 2 = 1$ is not equal to OV with $m \bmod 2 = 1$, whose OAM probability is $\delta_{(m \bmod 2 = 1)m'}$. Thus, MFV2 with $M \bmod 2 = 1$ cannot be generated by the rotationally symmetric superposition of two OVs with $m \bmod 2 = 1$, which is also the case of completely destructive interference [18, 20]. By using Eqs. (B6) and (B7), the unity summation of the OAM probability of $|M2\rangle$ can be proven, and the OAM mean of MFV2 for $M \in Z$ can be evaluated as $\sum_{m'=-\infty}^{\infty} m' P_{m'}(M2) = M$.

For $n = 4$,

$$\begin{aligned}\langle M4'|M4'\rangle &= \left\{\left[\langle M| + \langle M|\hat{U}^\dagger\left(\tfrac{\pi}{2}\right) + \langle M|\hat{U}^\dagger(\pi) + \langle M|\hat{U}^\dagger\left(\tfrac{3\pi}{2}\right)\right]\right. \\ &\quad \left.\left[|M\rangle + \hat{U}\left(\tfrac{\pi}{2}\right)|M\rangle + \hat{U}(\pi)|M\rangle + \hat{U}\left(\tfrac{3\pi}{2}\right)|M\rangle\right]\right\} \\ &= 4\left[\langle M|M\rangle + \operatorname{Re}\langle M|\hat{U}\left(\tfrac{\pi}{2}\right)|M\rangle + \operatorname{Re}\langle M|\hat{U}(\pi)|M\rangle + \operatorname{Re}\langle M|\hat{U}\left(\tfrac{3\pi}{2}\right)|M\rangle\right] \\ &= 4 + 4\cos(\pi M) + 8\cos^3\left(\tfrac{\pi}{2}M\right),\end{aligned} \quad (B8)$$

where the unitary property of $\hat{U}$, $e^{-i2\pi m} = 1$, and Eqs. (A3), (A8), and (A10) have been used.

$$\begin{aligned}
\left|\langle m'|M4'\rangle\right|^2 &= \left[\langle m'|M(\alpha)\rangle + \langle m'|\hat{U}\left(\tfrac{\pi}{2}\right)|M(\alpha)\rangle + \langle m'|\hat{U}(\pi)|M(\alpha)\rangle + \langle m'|\hat{U}\left(\tfrac{3\pi}{2}\right)|M(\alpha)\rangle\right] \\
&\quad \left[\langle m'|M(\alpha)\rangle + \langle m'|\hat{U}\left(\tfrac{\pi}{2}\right)|M(\alpha)\rangle + \langle m'|\hat{U}(\pi)|M(\alpha)\rangle + \langle m'|\hat{U}\left(\tfrac{3\pi}{2}\right)|M(\alpha)\rangle\right]^* \\
&= \left|\langle m'|M(\alpha)\rangle\right|^2 + \left|\langle m'|\hat{U}(\pi)|M(\alpha)\rangle\right|^2 + \left|\langle m'|\hat{U}\left(\tfrac{\pi}{2}\right)|M(\alpha)\rangle\right|^2 + \left|\langle m'|\hat{U}\left(\tfrac{3\pi}{2}\right)|M(\alpha)\rangle\right|^2 \\
&\quad + 2\operatorname{Re}\left[\langle m'|M(\alpha)\rangle\langle m'|\hat{U}\left(\tfrac{\pi}{2}\right)|M(\alpha)\rangle^*\right] + 2\operatorname{Re}\left[\langle m'|M(\alpha)\rangle\langle m'|\hat{U}(\pi)|M(\alpha)\rangle^*\right] \\
&\quad + 2\operatorname{Re}\left[\langle m'|M(\alpha)\rangle\langle m'|\hat{U}\left(\tfrac{3\pi}{2}\right)|M(\alpha)\rangle^*\right] + 2\operatorname{Re}\left[\langle m'|\hat{U}\left(\tfrac{\pi}{2}\right)|M(\alpha)\rangle\langle m'|\hat{U}(\pi)|M(\alpha)\rangle^*\right] \\
&\quad + 2\operatorname{Re}\left[\langle m'|\hat{U}(\pi)|M(\alpha)\rangle\langle m'|\hat{U}\left(\tfrac{3\pi}{2}\right)|M(\alpha)\rangle^*\right] + 2\operatorname{Re}\left[\langle m'|\hat{U}\left(\tfrac{\pi}{2}\right)|M(\alpha)\rangle\langle m'|\hat{U}\left(\tfrac{3\pi}{2}\right)|M(\alpha)\rangle^*\right] \\
&= \frac{4\sin^2(\pi\mu)}{\pi^2(M-m')^2}\left[1+\cos(\pi m') + 2\cos\left(\tfrac{\pi}{2}m'\right)\right],
\end{aligned}$$

(B9)

where Eqs. (A5), (A12), (A13), and $\cos(3m'\pi/2) = \cos(m'\pi/2)$ have been used. According to Eqs. (1), (B8), and (B9), the OAM probability distribution of MFV4 is

$$P_{m'}[M4(M)] = \frac{\left|\langle m'|M4'\rangle\right|^2}{\langle M4'|M4'\rangle} = \frac{4}{\pi^2}\sin^2\left(\frac{M\pi}{4}\right) \times \frac{1+\cos(\pi m') + 2\cos\left(\tfrac{\pi}{2}m'\right)}{(M-m')^2}, \quad \text{(B10)}$$

where $\sin^2(\mu\pi) = \sin^2(M\pi)$ has been used. The unity summation of the OAM probability of $|M4\rangle$ can be proven from Eq. (B10) as

$$\sum_{m'=-\infty}^{\infty} P_{m'}(M4) = \frac{4}{\pi^2}\sin^2\left(\frac{M\pi}{4}\right) \sum_{m'=-\infty}^{\infty} \frac{1+\cos(\pi m') + 2\cos\left(\tfrac{\pi}{2}m'\right)}{(M-m')^2} = 1, \quad \text{(B11)}$$

where Eqs. (C1), (C2), and (C5) have been used. The mean OAM of MFV2 can be calculated from Eq. (B10) as

$$\sum_{m'=-\infty}^{\infty} m' P_{m'}(M4) = \frac{4}{\pi^2}\sin^2\left(\frac{M\pi}{4}\right) \sum_{m'=-\infty}^{\infty} \frac{m' + m'\cos(\pi m') + 2m'\cos\left(\tfrac{\pi}{2}m'\right)}{(M-m')^2} = M - \frac{2}{\pi}\sin\left(M\frac{\pi}{2}\right), \quad \text{(B12)}$$

where Eqs. (C3), (C4), and (C6) have been used.

Let us consider the $M \in Z$ case in $n = 4$. As $M \bmod 4 = 0$, Eq. (B10) becomes

$$P_{m'}[M4(M \bmod 4 = 0)] = \begin{cases} 0, & m' \neq M \\ \dfrac{\sin^2(\mu\pi)}{\pi^2(M-m')^2} = 1, & m' = M \end{cases} = \delta_{Mm'}, \quad \text{(B13)}$$

where L'Hôpital's rule has been used for $m' = M$. According to Eq. (B13), MFV4 with $M \bmod 4 = 0$ is equal to OV with $m \bmod 4 = 0$. Thus, MFV4 with $M \bmod 4 = 0$ can be generated by the rotationally symmetric superposition of four OVs with $m \bmod 4 = 0$, the case of completely constructive interference [18, 20]. As $M \bmod 4 = 1, 2$, and $3$, Eq. (B10) respectively becomes

$$P_{m'}\left[M4(M \bmod 4 = 1)\right] = \frac{2\left[1+\cos(\pi m')+2\cos\left(\frac{\pi}{2}m'\right)\right]}{\pi^2(M-m')^2} = \begin{cases} \frac{2\left[1+\cos(\pi m')+2\cos\left(\frac{\pi}{2}m'\right)\right]}{\pi^2(M-m')^2}, & m' \bmod 4 = 0 \\ 0, & m' \bmod 4 = 1,\ 2,\ \text{and}\ 3 \end{cases},$$

(B14)

$$P_{m'}\left[M4(M \bmod 4 = 2)\right] = \frac{4\left[1+\cos(\pi m')+2\cos\left(\frac{\pi}{2}m'\right)\right]}{\pi^2(M-m')^2} = \begin{cases} \frac{4\left[1+\cos(\pi m')+2\cos\left(\frac{\pi}{2}m'\right)\right]}{\pi^2(M-m')^2}, & m' \bmod 4 = 0 \\ 0, & m' \bmod 4 = 1,\ 2,\ \text{and}\ 3 \end{cases},$$

(B15)

and

$$P_{m'}\left[M4(M \bmod 4 = 3)\right] = \frac{2\left[1+\cos(\pi m')+2\cos\left(\frac{\pi}{2}m'\right)\right]}{\pi^2(M-m')^2} = \begin{cases} \frac{2\left[1+\cos(\pi m')+2\cos\left(\frac{\pi}{2}m'\right)\right]}{\pi^2(M-m')^2}, & m' \bmod 4 = 0 \\ 0, & m' \bmod 4 = 1,\ 2,\ \text{and}\ 3 \end{cases},$$

(B16)

where the OAM modes of $m' \bmod 4 = 1,\ 2,$ and 3 vanish owing to the completely destructive interference in cascaded MZIs with $\theta = \pi$ and $\theta = \pi/2$ [18, 20]. According to Eqs. (B14), (B15), and (B16), MFV4 with $M \bmod 4 = 1,\ 2,$ and 3 is not equal to OV with $m \bmod 4 = 1,\ 2,$ and 3, whose OAM probability is $\delta_{(m \bmod 4 = 1,\ 2,\ \text{and}\ 3)m'}$. Thus, MFV4 with $M \bmod 4 = 1,\ 2,$ and 3 cannot be respectively generated by the rotationally symmetric superposition of four OVs with $m \bmod 4 = 1,\ 2,$ and 3, which are also the cases of completely destructive interference. By Eqs. (B13)–(B16), the unity summation of the OAM probability of $|M4\rangle$ can be proven, and the OAM mean of MFV4 in $M \in Z$ case can be evaluated as

$$\sum_{m'=-\infty}^{\infty} m' P_{m'}[M4] = \begin{cases} M, & M \bmod 4 = 0 \\ M - \frac{2}{\pi}, & M \bmod 4 = 1 \\ M, & M \bmod 4 = 2 \\ M + \frac{2}{\pi}, & M \bmod 4 = 3 \end{cases}. \quad \text{(B17)}$$

Note that in Eq. (B17), the OAM means for $M \bmod 4 = 1$ and 3 are not equal to those of the OVs with $m \bmod 4 = 1$ and 3, respectively. Alternatively, according to the principles of holography and light field phase shifting, MFV$n$ with $M \bmod n = 1,\ 2,\ ...,$ and $n-1$ can be generated, respectively, by using an $n$-fork hologram and $n$-section spiral phase plate [21].

## C. Formulas derived by using contour integration

According to the contour integral method [22],

$$\sum_{m'=-\infty}^{\infty} \frac{1}{(M-m')^2} = -\left[\text{sum of the residues of } \frac{\pi \cot(\pi m')}{(M-m')^2} \text{ at the poles of } \frac{1}{(M-m')^2}\right]$$

$$= -\lim_{m' \to M} \frac{d}{dm'}\left[(M-m')^2 \frac{\pi \cot(\pi m')}{(M-m')^2}\right] = \pi^2 \csc^2(M\pi),$$

(C1)

$$\sum_{m'=-\infty}^{\infty} \frac{\cos(\pi m')}{(M-m')^2} = -\left[\text{sum of the residues of } \frac{\pi \csc(\pi m')}{(M-m')^2} \text{ at the poles of } \frac{(-1)^{m'}}{(M-m')^2}\right]$$

$$= -\lim_{m' \to M} \frac{d}{dm'}\left[(M-m')^2 \frac{\pi \csc(\pi m')}{(M-m')^2}\right] = \pi^2 \csc(M\pi)\cot(M\pi),$$

(C2)

$$\sum_{m'=-\infty}^{\infty} \frac{m'}{(M-m')^2} = -\left[\text{sum of the residues of } \frac{\pi \cot(\pi m')m'}{(M-m')^2} \text{ at the poles of } \frac{m'}{(M-m')^2}\right]$$

$$= -\lim_{m' \to M} \frac{d}{dm'}\left[(M-m')^2 \frac{\pi m' \cot(\pi m')}{(M-m')^2}\right] = M\pi^2 \csc^2(M\pi) - \pi \cot(M\pi),$$

(C3)

and

$$\sum_{m'=-\infty}^{\infty} \frac{m' \cos(\pi m')}{(M-m')^2} = -\left[\text{sum of the residues of } \frac{\pi \csc(\pi m')m'}{(M-m')^2} \text{ at the poles of } \frac{(-1)^{m'} m'}{(M-m')^2}\right]$$

$$= -\lim_{m' \to M} \frac{d}{dm'}\left[(M-m')^2 \frac{\pi m' \csc(\pi m')}{(M-m')^2}\right] = \pi\left[(M\pi)\cot(M\pi) - 1\right]\csc(M\pi),$$

(C4)

Furthermore,

$$\sum_{m'=-\infty}^{\infty} \frac{\cos\left(\frac{\pi}{2}m'\right)}{(M-m')^2} = \sum_{m'=-\infty,\ldots,-2,0,2,4,\ldots} \frac{\cos\left(\frac{\pi}{2}m'\right)}{(M-m')^2} = \sum_{m''=-\infty}^{\infty} \frac{\cos(\pi m'')}{(M-2m'')^2}$$

$$= -\left[\text{sum of the residues of } \frac{\pi \csc(\pi m'')}{(M-2m'')^2} \text{ at the poles of } \frac{(-1)^{m''}}{(M-2m'')^2}\right]$$ (C5)

$$= -\lim_{m'' \to \frac{M}{2}} \frac{d}{dm''}\left[\left(m'' - \frac{M}{2}\right)^2 \frac{\pi \csc(\pi m'')}{(M-2m'')^2}\right] = \frac{\pi^2}{4} \csc\left(M\frac{\pi}{2}\right)\cot\left(M\frac{\pi}{2}\right)$$

and

$$\sum_{m'=-\infty}^{\infty} \frac{m' \cos\left(\frac{\pi}{2} m'\right)}{(M-m')^2} = \sum_{m'=-\infty,\ldots,-2,0,2,4,\ldots}^{\infty} \frac{m' \cos\left(\frac{\pi}{2} m'\right)}{(M-m')^2} = \sum_{m''=-\infty}^{\infty} \frac{2m'' \cos(\pi m'')}{(M-2m'')^2}$$

$$= -\left[\text{sum of the residues of } \frac{\pi \csc(\pi m'') 2m''}{(M-2m'')^2} \text{ at the poles of } \frac{(-1)^{m''} 2m''}{(M-2m'')^2}\right]$$

$$= -\lim_{m'' \to \frac{M}{2}} \frac{d}{dm''}\left[\left(m'' - \frac{M}{2}\right)^2 \frac{2\pi m'' \csc(\pi m'')}{(M-2m'')^2}\right] = \frac{\pi}{2}\left[\left(M\frac{\pi}{2}\right)\cot\left(M\frac{\pi}{2}\right) - 1\right]\csc\left(M\frac{\pi}{2}\right),$$

(C6)

where $\cos(m'\pi/2) = 0$, $m' = \pm 1, \pm 3, \ldots$ and $m'' = m'/2$ were used.

### D. Diagram of uncertainty relation between OAM and AP of MFV*n*

The OAM variance of an FV is divergent according to the formal uncertainty principle formula [11]. The OAM and AP uncertainties of various MFV*n* are compared by showing the respective OAM spectra and intensity distributions of the azimuthal coordinate in Fig. D1. The upper row of Fig. D1(a) presents near-field images of FVs with $M = 0.25891$ and $0.5$, for which the mean OAM $= 0.1$ and $0.5$, respectively. The profiles of their intensity gaps as indicated by the yellow lines in the top row are provided in the middle row of Fig. D1(a). The intensity distribution ranges in $\phi$, or AP uncertainties, can be compared by using either the low intensities of these images or the dips in the profiles. In the upper and middle rows of Fig. D1(a), the AP uncertainty for $M = 0.5$ is less than that for $M = 0.25891$. According to Eq. (A5), the OAM spectra of FVs with $M = 0.25891$ and $0.5$ are presented in the lower part of Fig. D1(a). As shown in the lower part of Fig. D1(a), the OAM spectral width, which indicates the OAM uncertainty, for $M = 0.5$ is greater than that for $M = 0.25891$. A small AP uncertainty implies a large OAM uncertainty, which shows the uncertainty relation of the OAM and AP for an FV.

The upper row of Fig. D1(b) depicts three near-field images of an FV, MFV2, and MFV4 with $M = 1/3, 2/3$, and $3/4$, respectively. The profiles of their intensity gaps as indicated by the yellow lines are shown with normalization in the middle row of Fig. D1(b). In the upper and middle rows of Fig. D1(b), the AP uncertainties are identical except for intermittent numbers of image profiles. The OAM spectra obtained using Eqs. (A5), (2), and (3) are presented in the lower row of Fig. D1(b). The OAM components vanish as $m' \bmod 1$, $m' \bmod 2$, and $m' \bmod 4 \neq 0$ for the FV, MFV2, and MFV4, respectively. Their OAM uncertainties are identical except for the resolutions of the OAM spectra, which are 1, 2, and 4 for the FV, MFV2, and MFV4, respectively. The resolution and intermittent number are inversely proportional. The simultaneously identical AP and OAM uncertainties demonstrate that the uncertainty relation of the AP and OAM is applicable to all MFV*n*.

The experimental setup shown in Fig. D1(c) was used to investigate the experimental beam profiles. The light beam was selected by an iris from the first-order diffraction of a hologram. Two lenses with focal lengths $f_1$ and $f_2$, respectively, were employed to control the beam size and propagation field conditions. The images were recorded using a CCD camera located at a distance $z$ from the conjugate plane of the hologram. The experimental data are presented in Figs. D1(d) and D1(e), which agree with the theoretical simulation data in Figs. D1(a) and D1(b), respectively.

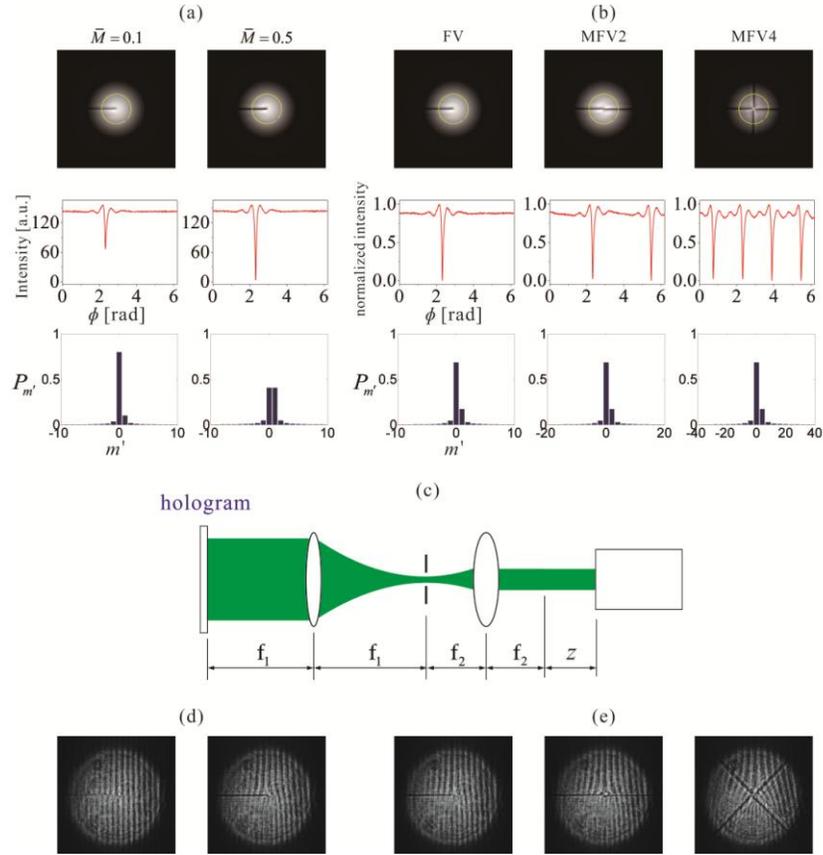

Fig. D1. Uncertainty relation for OAM and AP of MFV$n$. (a) Top: simulated intensity images of two FVs with mean OAMs of 0.1 and 0.5. Middle: intensity profiles along the yellow lines in the top figures. Bottom: OAM spectra for two FVs with mean OAMs of 0.1 and 0.5. (b) Top: three simulated intensity images of an FV with $M = 1/3$, MFV2 with $M = 2/3$, and MFV4 with $M = 4/3$. Middle: intensity profiles along the yellow lines in the top figures. Bottom: OAM spectra for the FV with $M = 1/3$, MFV2 with $M = 2/3$, and MFV4 with $M = 4/3$. (c) Experimental setup used to obtain images with propagation distance $z$. (d) Experimentally obtained images of two FVs with mean OAMs of 0.1 and 0.5. (e) Experimentally obtained images of an FV with $M = 1/3$, MFV2 with $M = 2/3$, and MFV4 with $M = 4/3$.